\def\lesssim{\mathrel{\hbox{\rlap{\hbox{\lower4pt\hbox{$\sim$}}}\hbox{$<$}}}}

\def\gtrsim{\mathrel{\hbox{\rlap{\hbox{\lower4pt\hbox{$\sim$}}}\hbox{$>$}}}}

\def\ll_lsun{log$({L/\rm L_{\odot}})$~}

\def\masa_msun{$M/ \rm M_{\odot}$~}

\def\m_mstar{$M/M_{*}$~}

\def\pg{\mbox{\object{PG~0122+200}}}
\def\pp{\mbox{\object{PG~1159$-$035}}}
\def\pr{\mbox{\object{PG~2131+066}}}
\def\pt{\mbox{\object{PG~1707+427}}}
\def\rxj{\mbox{\object{RX~J2117.1+3412}}}
\def\v4334{\mbox{\object{V4334~Sgr}}}

\documentclass{aa}
\usepackage{graphicx}
        
\begin{document}

\title{Asteroseismological constraints on the coolest GW Vir 
variable star (PG 1159-type) \pg}

\author{A. H. C\'orsico$^{1,2}$\thanks{Member of the Carrera del Investigador
Cient\'{\i}fico y Tecnol\'ogico, CONICET, Argentina.},
M. M. Miller Bertolami$^{1,2}$\thanks{Fellow of CONICET, Argentina.},
L. G. Althaus$^{1,2\star}$,
G. Vauclair$^{3}$,
\and K. Werner$^{4}$}

\offprints{A. H. C\'orsico}

\institute{
$^1$   Facultad   de   Ciencias  Astron\'omicas   y   Geof\'{\i}sicas,
Universidad  Nacional de  La Plata,  Paseo del  Bosque S/N,  (1900) La
Plata, Argentina.\\ 
$^2$ Instituto de Astrof\'{\i}sica La Plata, IALP,
CONICET-UNLP\\  
$^3$ Universite Paul Sabatier, Observatoire Midi-Pyrenees, CNRS/UMR5572,14 Av. Belin, 31400 Toulouse, France. \\  
$^4$ Institut f\"ur Astronomie und Astrophysik, Universit\"at
T\"ubingen, Sand 1, 72076 T\"ubingen, Germany.\\ 
\email{acorsico,althaus,mmiller@fcaglp.unlp.edu.ar; 
gerardv@ast.obs-mip.fr; werner@astro.uni-tuebingen.de} }

\date{Received; accepted}

\abstract{}{We present an asteroseismological study on 
\pg, the coolest known pulsating PG1159 (GW Vir)
star.  Our  results are based on  an augmented set of  the full PG1159
evolutionary models recently presented  by Miller Bertolami \& Althaus
(2006).}   {We perform  extensive computations  of  adiabatic $g$-mode
pulsation periods  on PG1159  evolutionary models with  stellar masses
ranging  from $0.530$ to  $0.741 M_{\odot}$.   These models  take into
account  the  complete  evolution  of progenitor  stars,  through  the
thermally pulsing AGB phase  and born-again episode.  We constrain the
stellar mass of \pg\ by comparing the observed period spacing with the
asymptotic period spacing and with  the average of the computed period
spacings. We also  employed the individual observed periods  to find a
representative seismological model for  \pg.}{We derive a stellar mass
of  $0.626 M_{\odot}$ from  a comparison  between the  observed period
spacing and the computed asymptotic period spacing, and a stellar mass
of $0.567 M_{\odot}$ by comparing the observed period spacing with the
average of the computed period spacing.  We also find, on the basis of
a period-fit procedure, an asteroseismological model representative of
\pg\ which  is able to reproduce  the observed period  pattern with an
average of the period differences of $\overline{\delta
\Pi_i}= 0.88$  s and a root-mean-square  residual of $\sigma_{_{\delta
\Pi_i}}=  1.27$ s.   The model  has an  effective  temperature $T_{\rm
eff}= 81\,500$  K, a  stellar mass $M_*=  0.556 M_{\odot}$,  a surface
gravity  $\log   g=  7.65$,  a   stellar  luminosity  and   radius  of
$\log(L_*/L_{\odot})   =  1.14$   and   $\log(R_*/R_{\odot})=  -1.73$,
respectively, and a He-rich envelope thickness of $M_{\rm env}= 1.9
\times 10^{-2} M_{\odot}$.  We derive a seismic distance $d \sim 614$
pc and a  parallax $\pi \sim 1.6$ mas.  The  results of the period-fit
analysis carried out in this work suggest that the asteroseismological
mass of \pg\ could be $\sim 6 - 20 \%$ lower than thought hitherto and
in closer  agreement (to  within $\sim 5  \%$) with  the spectroscopic
mass. This  result suggests that a reasonable  consistency between the
stellar  mass values obtained  from spectroscopy  and asteroseismology
can  be   expected  when  detailed  PG1159   evolutionary  models  are
considered.}{}
  

\authorrunning{C\'orsico et al.}

\titlerunning{Asteroseismological constraints on \pg}

\maketitle

 
\section{Introduction}
\label{intro}

\pg\ (BB Psc or WD 0122+200) is the coolest known pulsating PG1159 
star belonging to the GW Vir class of variables. GW Vir stars are very
hot hydrogen-deficient  post-Asymptotic Giant Branch  (AGB) stars with
surface layers  rich in  helium, carbon and  oxygen (Werner  \& Herwig
2006).  They exhibit  multiperiodic luminosity variations with periods
in  the  range $5-50$  minutes,  attributable  to nonradial  pulsation
$g$-modes.   PG1159 stars  are  thought to  be  the evolutionary  link
between Wolf-Rayet type central stars of planetary nebulae and most of
the  hydrogen-deficient white  dwarfs. It  is generally  accepted that
these stars  have their  origin in a  born-again episode induced  by a
post-AGB  helium  thermal pulse  (see  Iben  et  al. 1983,  Herwig  et
al.  1999, Lawlor  \&  MacDonald  2003, Althaus  et  al. 2005,  Miller
Bertolami et al. 2006 for recent references).

\pg\ is characterized by $T_{\rm  eff}= 80\, 000 \pm 4\, 000$~K 
and $\log g= 7.5 \pm 0.5$  (Dreizler \& Heber 1998). At this effective
temperature, \pg\  currently defines  the locus of  the low-luminosity
red edge of the GW  Vir instability strip.  The photometric variations
of this  star were discovered by  Bond \& Grauer  (1987).  Besides the
intrinsic interest in probing its interior, pulsation studies of
\pg\ offer a
unique  opportunity  to  study   neutrino  physics.   Indeed,  at  the
evolutionary stage characterizing \pg, neutrino emission constitutes a
main  energy   sink\footnote{At  variance  with   the  solar  neutrino
emission, which is  a by-product of nuclear fusion,  the neutrino flux
of  pre-white dwarf  stars such  as \pg\  is the  result  of different
scattering   processes,   being   the   {\it   plasmaneutrino},   {\it
Bremsstrahlung  neutrino} and  {\it photoneutrino}  emission  the most
relevant ones (see O'Brien \& Kawaler 2000).}  (O'Brien et al.  1998).

The determination of the stellar mass  of \pg\ has been the subject of
numerous  investigations.   The stellar  mass  of pulsating  pre-white
dwarfs can be constrained,  in principle, from asteroseismology ---the
asteroseismological  mass--- either through  the {\it  observed period
spacing}  (see, for instance,  Kawaler \&  Bradley 1994;  C\'orsico \&
Althaus 2006)  or by  means of the  {\it individual  observed periods}
(see,  e.g.,  Kawaler \&  Bradley  1994,  C\'orsico  \& Althaus  2006,
C\'orsico et  al. 2007).   The early study  of O'Brien et  al.  (1996)
predicts  a  stellar mass  of  about  $0.66-0.72  M_{\odot}$ for  \pg\
corresponding to an observed mean  period spacing of 21.2 s.  Vauclair
et al. (1995), on the other hand, suggest an even higher stellar mass,
based on a observed mean period spacing of \pg\ of $\sim 16$ s. Later,
O'Brien et al. (1998) find strong evidence for a $\ell= 1$ mean period
spacing  of  21  s,  although  a  value  of  $\sim  16$  s  cannot  be
conclusively ruled  out.  These values  of the period spacing  imply a
stellar  mass  of $\sim  0.69  M_{\odot}$  and  $\sim 1.0  M_{\odot}$,
respectively, based on the PG 1159 models then available.  By means of
a  period-fit  procedure  based  on PG1159  evolutionary  models  with
several masses derived from the  full sequence of $0.589 M_{\odot}$ of
Althaus et  al. (2005), C\'orsico  \& Althaus (2006) obtain  a stellar
mass  of $M_*= 0.64  M_{\odot}$ for  \pg. Recently,  Fu et  al. (2007)
(hereinafter   FUEA07)  have   presented  new   multisite  photometric
observations of \pg\ obtained in 2001 and 2002.  By collecting the new
data together with previous observations, these authors have succeeded
in detecting  a total  of 23 frequencies  corresponding to  modes with
$\ell= 1$ and  derived unambiguously a mean period  spacing of 22.9~s.
On the basis of the models of Kawaler \& Bradley (1994), these authors
inferred a stellar mass of $0.59 \pm 0.02 M_{\odot}$.

The stellar  mass of  PG1159 stars can  also be estimated  through the
comparison of the  spectroscopic values of $T_{\rm eff}$  and $g$ with
evolutionary tracks  ---the spectroscopic mass.   On the basis  of the
evolutionary tracks  of O'Brien \&  Kawaler (2000), Dreizler  \& Heber
(1998) derived a stellar mass of  $0.53 \pm 0.1 M_{\odot}$ for \pg. On
the  other  hand,  Werner  \&  Herwig  (2006)  determined  $M_*=  0.58
M_{\odot}$ from  a comparison with  the H-rich evolutionary  models of
Sch\"onberner (1983).  The most recent determination is that of Miller
Bertolami  \& Althaus  (2006), who  derived  a stellar  mass of  $0.53
M_{\odot}$ on the basis of  PG1159 evolutionary models that take fully
into account  the evolutionary history and the  surface composition of
the progenitor stars.

The discrepancy between the asteroseismological mass derived by FUEA07
($0.57   \leq  M_*/M_{\odot}   \leq   0.61$)  and   the  most   recent
spectroscopic  determination  ($0.53 M_{\odot}$)  has  prompted us  to
undertake  the  present  asteroseismological  investigation  for  \pg,
taking  full advantage of  the new  generation of  PG1159 evolutionary
models  recently  developed by  Miller  Bertolami  \& Althaus  (2006).
These authors have  followed in detail all of  the evolutionary phases
prior to the formation of  PG1159 stars with different stellar masses,
particularly the born-again stage. In addition to the
issue  of the  stellar mass,  the employment  of such  detailed PG1159
models allows us to address  the question of the He-rich envelope mass
($q_y \equiv M_{\rm env}/M_{\odot}$) of \pg, which FUEA07 
constrain to be in the range $-6
\lesssim \log q_y  \lesssim -5.3$.  Finally, a  precise knowledge of 
the mass  of \pg\ is  a crucial aspect  concerning the role  played by
neutrinos in that star.

The  paper is organized  as follows:  in the  next Section  we briefly
describe      our       PG1159      evolutionary      models.       In
Sect.~\ref{period-spacing} we derive the stellar mass of \pg\ by means
of  the observed  period  spacing.  In  Sect.~\ref{fitting} we  derive
structural  parameters  of  this  star  by  employing  the  individual
observed periods.   In this  section we derive  an asteroseismological
model representative of \pg\ (\S~\ref{searching}) and discuss its main
structural and pulsational characteristics (\S~\ref{char}), its helium
envelope  thickness  (\S~\ref{helium}),  its mode-trapping  properties
(\S~\ref{mode-trapping})  and  the  asteroseismological  distance  and
parallax (\S~\ref{distance}).   Finally, in Sect.~\ref{conclusions} we
summarize our main results and make some concluding remarks.

\begin{figure}
\centering
\includegraphics[clip,width=240pt]{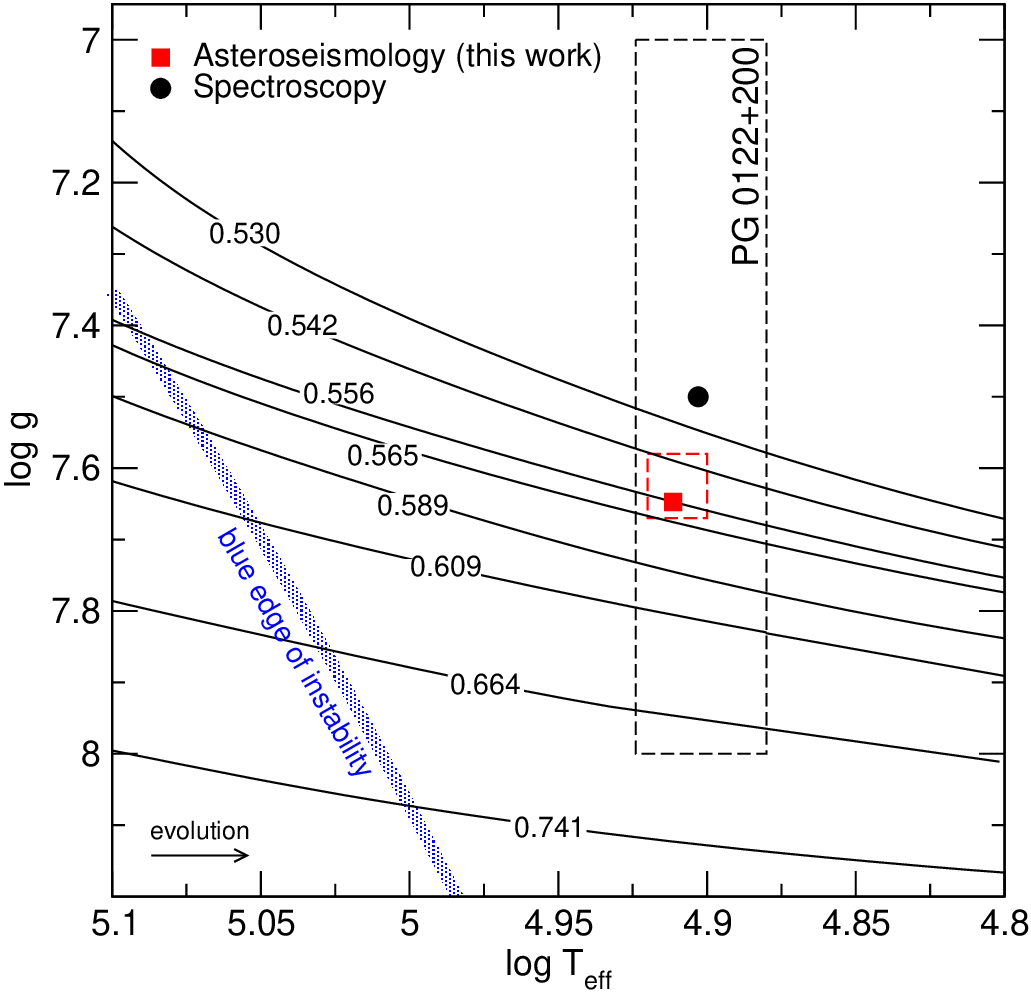}
\caption{Our PG1159 full evolutionary  tracks in  
the $\log T_{\rm eff} -\log  g$ plane, labelled with the corresponding
stellar mass value  in solar units.  The black  circle is the location
of  \pg\ according to  spectroscopy ($T_{\rm  eff}= 80  \pm 4$  kK and
$\log  g=  7.5  \pm 0.5$).  Note  the  large  error box  (dashed),  in
particular for $\log g$.  The square (red) is the location of the star
as predicted by our asteroseismological analysis (see \S
\ref{fitting}).  The blue  (hot)  
boundary  of the  theoretical  dipole ($\ell=  1$) instability  domain
---according  to C\'orsico  et  al.  (2006)---  is  also shown  [Color
figure only available in the electronic version of the article.] }
\label{figure1}
\end{figure}

\section{Evolutionary models and numerical tools}
\label{evolutionary}

The  pulsation  analysis  presented  in  this work  relies  on  a  new
generation  of stellar  models  that take  into  account the  complete
evolution of PG1159 progenitor  stars. These models have been recently
employed by our group for a pulsation stability analysis of the GW Vir
stars  and for  an asteroseismological  study of  the hot  PG1159 star
\rxj\   (C\'orsico  et   al.  2006   and  C\'orsico   et   al.   2007,
respectively).  The  stages for the formation and  evolution of PG1159
stars  were  computed with  the  LPCODE  evolutionary  code, which  is
described at length in Althaus et al.  (2005). The neutrino production
rates  adopted in our  computations are  those of  Itoh et  al. (1989,
1992).

Specifically, the background of  stellar models was extracted from the
evolutionary calculations recently presented by Althaus et al. (2005),
Miller Bertolami \&  Althaus (2006), and C\'orsico et  al. (2006), who
computed the  complete evolution of model star  sequences with initial
masses on  the ZAMS in the range  $1 - 3.75 M_{\odot}$.   We refer the
reader to those  works for details. Suffice it to  mention that all of
the  post-AGB evolutionary  sequences have  been followed  through the
very late  thermal pulse (VLTP)  and the resulting  born-again episode
that give  rise to the  H-deficient, helium-, carbon-  and oxygen-rich
composition  characteristic  of  PG1159  stars.   The  masses  of  the
resulting  remnants  are 0.530,  0.542,  0.556,  0.565, 0.589,  0.609,
0.664,  and  $0.741  M_{\odot}$.   The  new sequence  with  $M=  0.556
M_{\odot}$, coming from a progenitor star with $M_*= 1.8 M_{\odot}$ at
the   ZAMS,   has  been   computed   specifically   for  the   present
asteroseismological  study.   The  evolutionary  tracks in  the  $\log
T_{\rm eff}  - \log g$  plane for the  PG1159 regime are  displayed in
Fig. \ref{figure1}.

\begin{figure}
\centering
\includegraphics[clip,width=240pt]{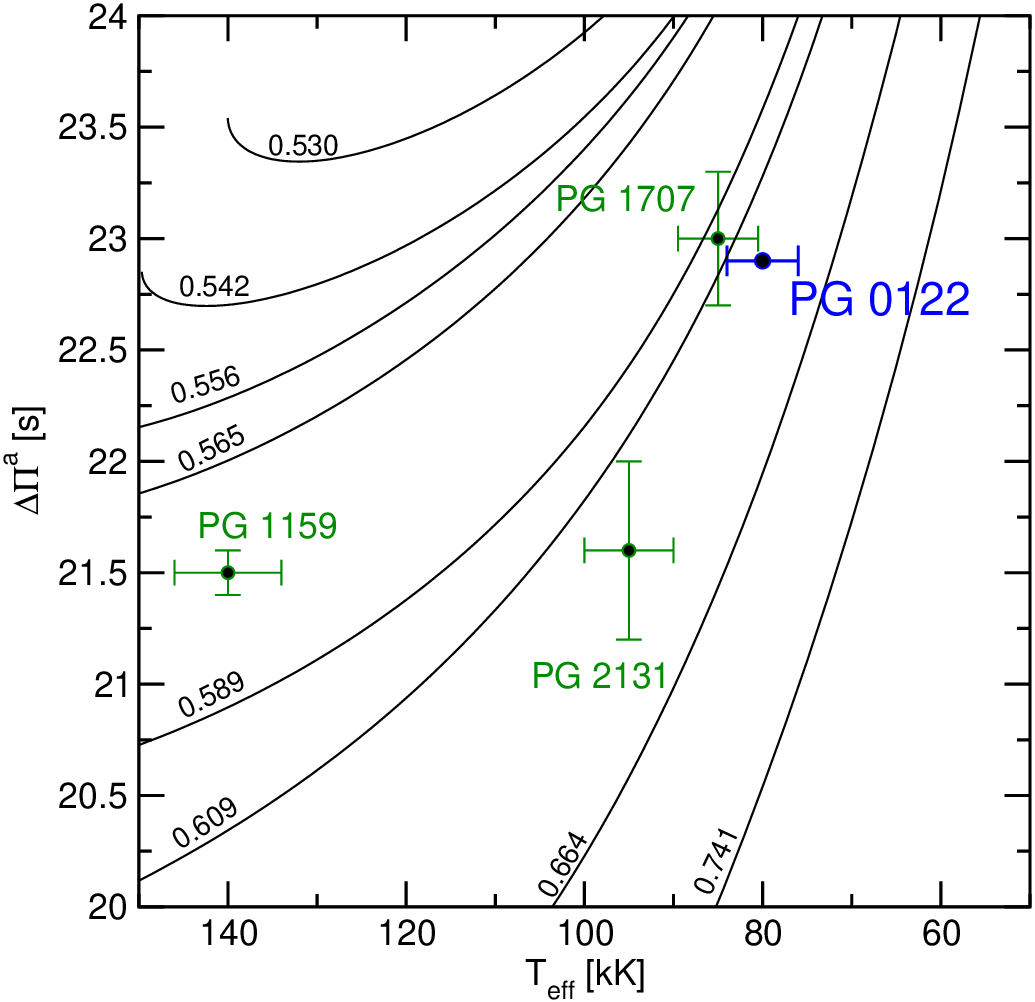}
\caption{The dipole ($\ell= 1$) asymptotic period spacing 
($\Delta \Pi_{\ell}^{\rm a}$) for different stellar masses in terms of
the  effective  temperature.   Numbers  along each  curve  denote  the
stellar masses (in  solar units). The plot also  shows the location of
\pg\ ($T_{\rm eff}= 80 \pm 4$ kK and $\Delta \Pi^{\rm O}= 22.9$~s) and
the remainder high-gravity, low-luminosity GW Vir stars (\pp, \pr, and
\pt) with the period spacing and $T_{\rm eff}$
data taken from Kawaler et al.  (2004). The mass of \pg\ as derived by
comparing  $\Delta \Pi_{\ell}^{\rm  a}$ with  $\Delta \Pi^{\rm  O}$ is
$M_*= 0.625_{-0.016}^{+0.019} M_{\odot}$  [Color figure only available
in the electronic version of the article.]. }
\label{figure2}
\end{figure}

It  is worth  mentioning that  the  use of  these evolutionary  tracks
constitutes   a   major   improvement   with   respect   to   previous
asteroseismological  studies.  As  mentioned, our  PG1159 evolutionary
sequences  are  derived  from  the complete  born-again  evolution  of
progenitor stars  and a careful  treatment of the  mixing and extramixing 
processes during the core helium  burning, fundamental aspects when attempts are
made at constructing stellar  models appropriate for PG1159 stars.  In
particular, these  evolutionary calculations reproduce:  (1) the spread
in surface chemical composition observed  in PG1159 stars, (2) the
short born-again  times of \v4334\  (see Miller Bertolami et  al. 2006
and Miller Bertolami \& Althaus 2007a), (3) the location of the GW Vir 
instability strip in the $\log T_{\rm eff}- \log g$ plane 
(C\'orsico et al. 2006), and (4) the expansion age of the 
planetary nebula of \rxj\ (see the paper by C\'orsico et al. 2007 and 
its associated erratum). We believe that the employment
of  these  new evolutionary  computations  render  reliability to  our
pulsational inferences for \pg.

We  computed  $\ell=  1$  $g$-mode  adiabatic  pulsation  periods  and
asymptotic period spacings with the same numerical code we employed in
our previous works (see, e.g., C\'orsico \& Althaus 2006 for details).
We  analyzed  about  3000  PG1159  models covering  a  wide  range  of
effective  temperatures  and  luminosities ($5.4  \gtrsim  \log(T_{\rm
eff}) \gtrsim 4.8$ and  $0 \lesssim \log(L_*/L_{\odot}) \lesssim 4.2$,
respectivley) and a range of stellar masses ($0.530 \leq M_*/M_{\odot}
\leq 0.741$). 

\begin{figure}
\centering
\includegraphics[clip,width=240pt]{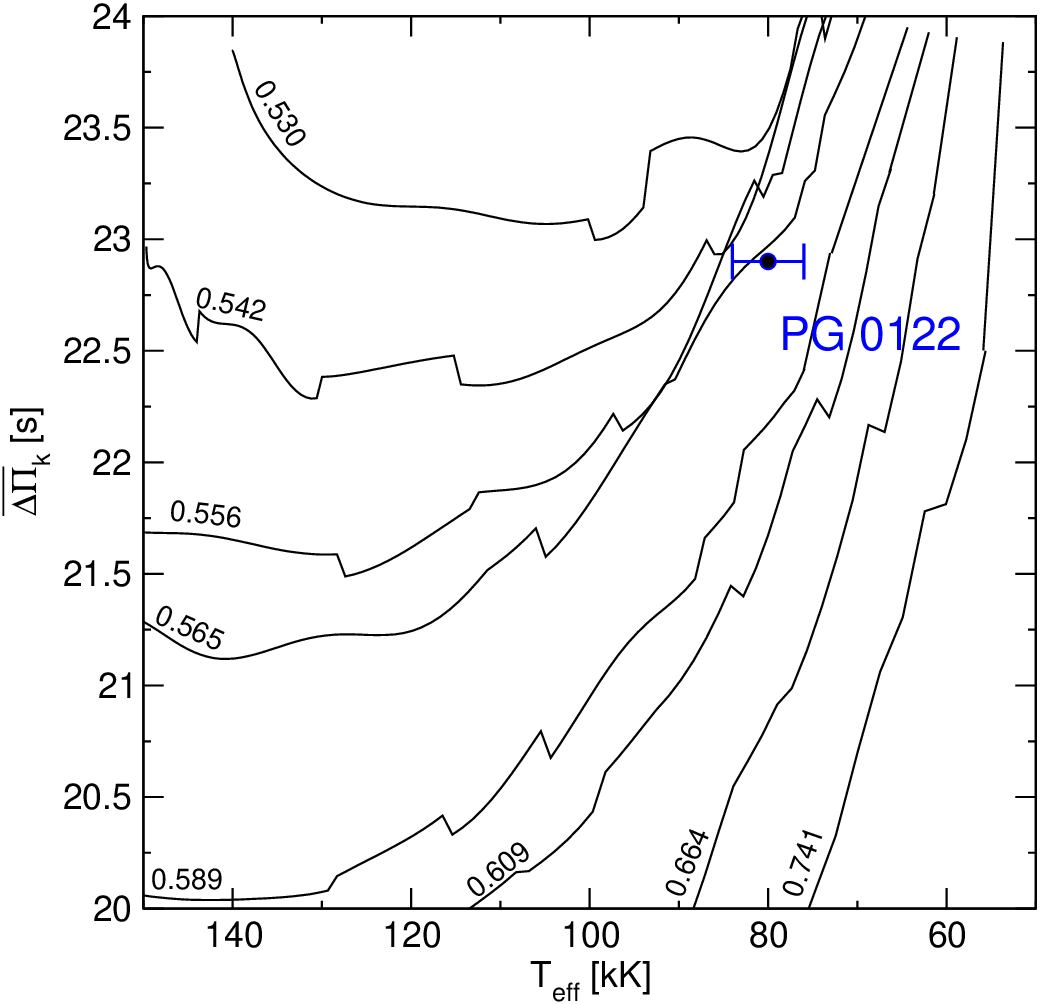}
\caption{Same as Fig. \ref{figure2}, but for the average 
of  the computed period  spacings ($\overline{\Delta  \Pi_{k}}$).  The
 mass of \pg\ as derived by comparing $\overline{\Delta \Pi_{k}}$ with
 $\Delta  \Pi^{\rm  O}$  is $M_*=  0.567_{-0.013}^{+0.007}  M_{\odot}$
 [Color  figure  only  available  in  the electronic  version  of  the
 article].}
\label{figure3}
\end{figure}

\section{Mass determination from the observed period spacing}
\label{period-spacing}

Here,  we  constrain  the  stellar  mass  of  \pg\  by  comparing  the
asymptotic  period  spacing,  $\Delta  \Pi_{\ell}^{\rm  a}$,  and  the
average of  the computed period  spacings, $\overline{\Delta \Pi_{k}}$
($k$ being the radial order),  with the {\it observed} period spacing,
$\Delta  \Pi^{\rm  O}$.\footnote{Note  that  most  asteroseismological
studies rely  on the  asymptotic period spacing  to infer  the stellar
mass of  GW Vir pulsators.} These  methods take full  advantage of the
fact that the period spacing  of PG1159 pulsators depends primarily on
the  stellar  mass, and  weakly  on  the  luminosity and  the  He-rich
envelope mass fraction (Kawaler  \& Bradley 1994; C\'orsico \& Althaus
2006).  Note that in these  approachs we make use of the spectroscopic
constraint  that the  effective  temperature  of \pg\  is  of $80$  kK
(Dreizler \& Heber 1998).

The asymptotic period  spacing and the average of  the computed period
spacings  for  $\ell=  1$  modes   as  a  function  of  the  effective
temperature are  displayed in Figs.   \ref{figure2} and \ref{figure3},
respectively,  for  different  stellar  masses. Also  shown  in  these
diagrams is  the location  of \pg, with  $\Delta \Pi^{\rm  O}= 22.9$~s
(FUEA07). Here, $\Delta \Pi_{\ell}^{\rm a}=
\Pi_0 / \sqrt{\ell(\ell+1)}$, where $\Pi_0= 2 \pi^2 [ \int_{r_1}^{r_2}
(N/r) dr]^{-1}$, being  $N$ the Brunt-V\"ais\"al\"a frequency (Tassoul
et al. 1990).  The  quantity $\overline{\Delta \Pi_{k}}$, on the other
hand, is  assessed by averaging  the computed forward  period spacings
($\Delta \Pi_{k}=  \Pi_{k+1}- \Pi_{k}$) in  the range of  the observed
periods in \pg\ (330-620 s; see Table \ref{tabla1}).

From a comparison between $\Delta \Pi^{\rm O}$ and $\Delta
\Pi_{\ell}^{\rm a}$ we obtain a stellar mass of
$M_*= 0.625_{-0.016}^{+0.019} M_{\odot}$.  The quoted uncertainties in
the  value  of  $M_*$  come  from  the  errors  in  the  spectroscopic
determination of the  effective temperature.  In the same  way, we get
$M_*= 0.567_{-0.013}^{+0.007} M_{\odot}$ if we compare $\Delta
\Pi^{\rm O}$ and $\overline{\Delta \Pi_{k}}$. 
The  higher value  of $M_*$  (about $10  \%$ larger)  as  derived from
$\Delta  \Pi_{\ell}^{\rm a}$  is due  to that  usually  the asymptotic
period  spacing is  larger than  the  average of  the computed  period
spacings (see C\'orsico \& Althaus  2006), in particular for the short
periods  like  those  exhibited  by  \pg, i.e.   for  which  the  full
asymptotic   regime  of   the  modes   ($k  \gg   1$)  has   not  been
attained\footnote{At  variance  with  this,  for  the  longer  periods
exhibited  by  \rxj\  (with  $30\leq  k  \leq  53  $)  the  asymptotic
conditions are more nearly reached  and, as a result, the stellar mass
derived  from the  asymptotic period  spacing  is very  close to  that
derived from de average of the computed period spacings (see C\'orsico
et al.   2007).}.  It is  important to note  that the first  method to
derive the  stellar mass  is somewhat less  realistic than  the second
one, because the asymptotic  predictions are, in principle, only valid
for chemically homogeneous stellar models, while our PG1159 models are
indeed chemically stratified.

Finally, we note that our inferred stellar mass values of $M_* \approx
0.57 M_{\odot}$ and in particular  $M_* \approx 0.63 M_{\odot}$ are in
conflict  with  the  value  $M_*=  0.53  M_{\odot}$  as  derived  from
spectroscopy coupled  to evolutionary tracks (Dreizler  \& Heber 1998;
Miller Bertolami \& Althaus 2006).

\section{Constraints from the individual observed periods}
\label{fitting}

\subsection{The search for the best-fit model}
\label{searching}

\begin{figure}
\centering
\includegraphics[clip,width=250pt]{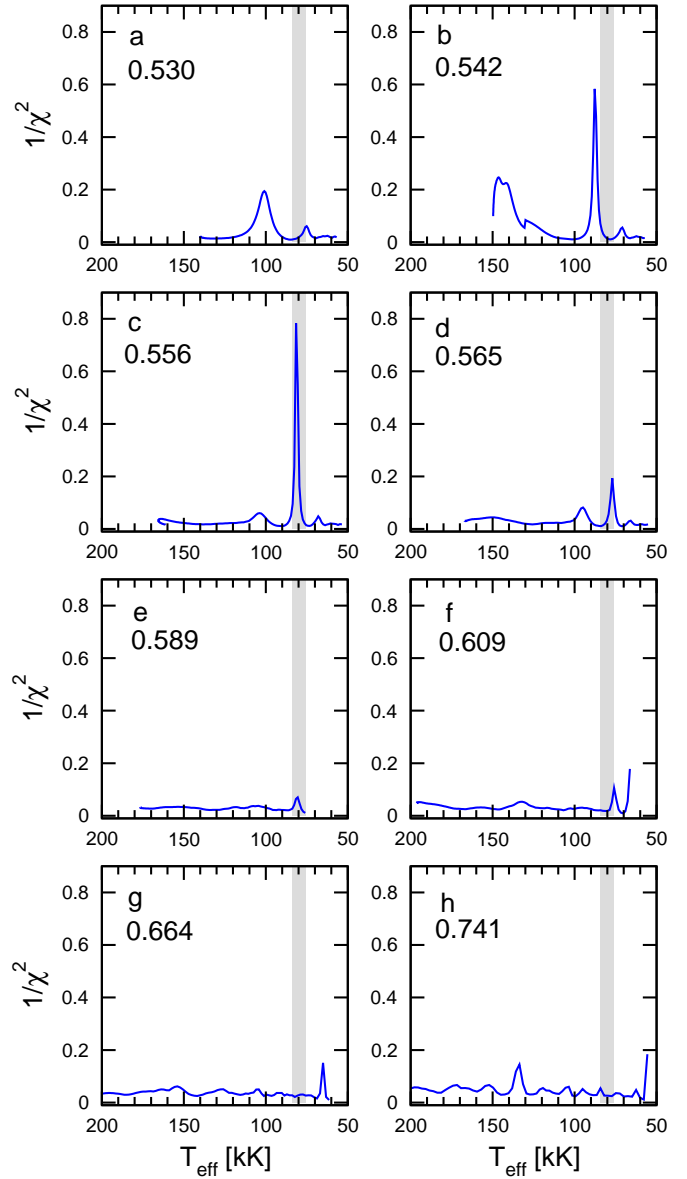}
\caption{The inverse of the quality function of the period fit 
in terms  of the effective  temperature for the PG1159  sequences with
different stellar masses indicated (in solar mass) in each panel.  The
grey  vertical  strip   corresponds  to  the  spectroscopic  effective
temperature of  \pg\ and  its uncertainties ($T_{\rm  eff}= 80\,000\pm
4\,000$  K).  Note  the strong  maximum in  $(\chi^2)^{-1}$  for $M_*=
0.556 M_{\odot}$ at $T_{\rm  eff} \approx 81\,500$ K. This corresponds
to our  ``best-fit'' model (see  text for details) [Color  figure only
available in the electronic version of the article].}
\label{figure4} 
\end{figure}

In this approach we seek a pulsation model that best matches the
\emph{individual}  pulsation  periods of  \pg.   We assume that all of the 
observed  periods correspond  to $\ell=  1$ modes  (see  FUEA07).  The
goodness  of  the  match  between the  theoretical  pulsation  periods
($\Pi_k$)  and the  observed individual  periods ($\Pi_i^{\rm  O}$) is
measured by means of a quality function defined as $\chi^2(M_*, T_{\rm
eff})=
\sum_{i=1}^{n} \min[(\Pi_i^{\rm O}- \Pi_k)^2]/n$, where 
$n$ (= 9) is the number of observed periods (first column in Table
\ref{tabla1}).  The PG 1159 model that shows the lowest value of
$\chi^2$ will be adopted as  the ``best-fit model''. This approach has
also  been  used by  C\'orsico  \&  Althaus  (2006) and  C\'orsico  et
al. (2007).

We evaluate the function $\chi^2(M_*, T_{\rm eff})$ for stellar masses
of  $0.530, 0.542,  0.556,  0.565, 0.589,  0.609,  0.664$, and  $0.741
M_{\odot}$.   For the effective  temperature we  employed a  much more
finer   grid  ($\Delta   T_{\rm   eff}=  10-30$   K).   The   quantity
$(\chi^2)^{-1}$ in  terms of  the effective temperature  for different
stellar masses is  shown in the mosaic of  Fig. \ref{figure4} together
with  the spectroscopic  effective temperature  of \pg.   We  find one
strong maximum for  the model with $M_*= 0.556  M_{\odot}$ and $T_{\rm
eff}\approx 81.5$ kK (panel {\bf c}). Such a pronounced maximum in the
inverse  of  $\chi^2$  implies  an  excellent  agreement  between  the
theoretical  and observed  periods. Another  maximum,  albeit somewhat
less  pronounced,  is  encountered  for  the model  with  $M_*=  0.542
M_{\odot}$  at $T_{\rm  eff}\approx 87.7$  kK and  constitutes another
acceptable  asteroseismological solution,  in  particular because  its
stellar mass is closer to the spectroscopic mass of
\pg\ ($0.53 M_{\odot}$).
However,  because  the  agreement  between  observed  and  theoretical
periods for this model is somewhat  poorer than for the one with $M_*=
0.556 M_{\odot}$, we  choose to adopt this last  model as the best-fit
asteroseismological  model.   Note  that  our best-fit  model  has  an
effective temperature  very close  to that suggested  by spectroscopy,
well inside the error bar.   A detailed comparison of the observed $m=
0$ periods in
\pg\  with the theoretical  periods of the best-fit  model is
provided in Table \ref{tabla1}.  The high quality of our period fit is
quantitatively  reflected  by  the  average  of  the  absolute  period
differences, $\overline{\delta \Pi_i}= (\sum_{i=1}^n |\delta
\Pi_i|)/n$, where $\delta \Pi_i= \Pi_i^{\rm O} -\Pi_k$, and by the
root-mean-square  residual,   $\sigma_{_{\delta  \Pi_i}}=  \sqrt{(\sum
|\delta \Pi_i|^2)/n}  $.  We obtain $\overline{\delta  \Pi_i}= 0.88$ s
and $\sigma_{_{\delta \Pi_i}}= 1.27$ s.  The quality of our fit for
\pg\ is much better than that  achieved by C\'orsico et al. (2007) for
\rxj\  ($\overline{\delta \Pi_i}= 1.08$ s)  and those obtained
by  Kawaler  \&  Bradley   (1994)  and  C\'orsico  \&  Althaus  (2006)
($\overline{\delta
\Pi_i}= 1.19$ s and $\overline{\delta \Pi_i}= 1.79$ s, respectively) 
for  \pp. Note that  we are  able to  get a  PG1159 model  that nicely
reproduces the period spectrum observed in \pg\
\emph{without artificially tuning} the value of structural parameters such as
the thickness of the  outer envelope, the surface chemical abundances,
or the  shape of  the core chemical  profile which, instead,  are kept
fixed at the values predicted by the evolutionary computations.
  
\begin{table}
\centering
\caption{Observed $m= 0$ periods ($\Pi_i^{\rm O}$) for \pg\ 
(taken  from FUEA07), theoretical  $\ell=  1, m= 0$  
periods ($\Pi_k$), period   differences  
($\delta   \Pi_i= \Pi_i^{\rm O}- \Pi_k$),  radial orders ($k$),  
linear  growth   rates ($\eta_k$), and rates of 
period change ($\dot{\Pi}_k$) for the best-fit model.}
\begin{tabular}{cccccc}
\hline
\hline
$\Pi_i^{\rm O}$ & $\Pi_k$ & $\delta \Pi_i$ & $k$ & $\eta_k$ & 
$\dot{\Pi}_k$\\
$[$s$]$ & $[$s$]$ & $[$s$]$ & & $[10^{-6}]$& $[10^{-12}$ s/s$]$\\
\hline
$336.68$ & $334.12$ & $2.56$   & 12 & $1.12$  & 1.22\\
---      & $354.85$ & ---      & 13 & $1.74$  & 1.70\\
$380.10$ & $380.70$ & $-0.60 $ & 14 & $3.54$  & 1.84\\
$400.99$ & $400.35$ & $0.64$   & 15 & $6.13$  & 1.59\\
---      & $425.25$ & ---      & 16 & $8.53$  & 2.24\\
$449.48$ & $448.16$ & $1.32$   & 17 & $15.90$ & 1.71\\
$468.69$ & $469.32$ & $-0.63$  & 18 & $16.30$ & 2.46\\
$494.92$ & $494.76$ & $0.16$   & 19 & $28.64$ & 2.19\\
$517.96$ & $516.65$ & $1.31$   & 20 & $36.50$ & 2.10\\
---      & $539.65$ & ---      & 21 & $32.24$ & 3.11\\
$564.28$ & $563.98$ & $0.30$   & 22 & $61.06$ & 2.00\\
---      & $585.80$ & ---      & 23 & $48.17$ & 3.04\\
$611.15$ & $610.73$ & $0.42$   & 24 & $55.99$ & 3.26\\
\hline
\hline
\end{tabular}
\label{tabla1}
\end{table}

Table \ref{tabla1}  also shows the  linear growth rates  ($\eta_k$) of
the  fitted pulsation  modes (fifth  column) for  our  best-fit model,
computed with  the nonadiabatic pulsation code  described in C\'orsico
et al.  (2006).   We found that all of the  fitted modes have positive
values  of $\eta_k$,  implying pulsational  instability,  although our
stability analysis predicts a band of unstable mode-periods ($230
\lesssim \Pi_k \lesssim  730 $ s) somewhat wider  than the interval of
periods detected in \pg.

The last column in Table  \ref{tabla1} shows the rate of period change
of the  fitted pulsation  modes. Our calculations  predict all  of the
pulsation periods  to \emph{increase} with time  ($\dot{\Pi}_k>0$), in
accordance with  the decrease of the  Brunt-V\"ais\"al\"a frequency in
the core of  the model induced by cooling. Note  that at the effective
temperature of  \pg, cooling has the largest  effect on $\dot{\Pi}_k$,
while  gravitational  contraction,  which  should  result  in  a  
\emph{decrease} of periods  with  time, becomes  negligible  and no  longer
affects  the pulsation  periods, except  for  the case  of modes  
\emph{trapped} in the  envelope (see \S \ref{mode-trapping}).  
Until now, the only  secure measurement of
$\dot{\Pi}$ in pre-white dwarf stars  is that of \pp, the prototype of
the class, for which Costa et al.  (1999) obtained a positive value of
$\dot{\Pi}=  (+1.307 \pm  0.003) \times  10^{-10}$ s/s  for the  516 s
period. Note that our  theoretical (positive) $\dot{\Pi}_k$ values for
the best-fit model ($1.22-3.26 \times 10^{-12}$ s/s) are two orders of
magnitude  lower.   For  the  case  of \pg,  a  determination  of  any
$\dot{\Pi}$ has  not been  assessed up to  now, although work  in this
direction is in progress (see Fu et al. 2002).

\begin{table*}
\centering
\caption{The main characteristics of \pg. The second column  
corresponds to spectroscopic results, whereas the third and fourth 
columns present results from the pulsation study of FUEA07
and from the asteroseismological model of this work, respectively.} 
\begin{tabular}{l|ccc}
\hline
\hline
Quantity                     & Spectroscopy                     & FUEA07                  & Asteroseismology             \\
                             &                                  &                         & (This work)                  \\  
\hline
$T_{\rm eff}$ [kK]           & $80 \pm 4^{\rm (a)}$             &        ---              & $81.54_{-1.4}^{+0.8}$        \\
$M_*$ [$M_{\odot}$]          & $0.53\pm 0.1^{\rm (b)}$          & $0.59 \pm 0.02$         & $0.556_{-0.014}^{+0.009}$    \\ 
$\log g$ [cm/s$^2$]          & $7.5 \pm 0.5^{\rm (a)}$          &        ---              & $7.65_{-0.07}^{+0.02} $      \\ 
$\log (L_*/L_{\odot})$       & ${1.2_{-0.3}^{+0.2}}^{(**)}$     & $1.3 \pm 0.5$           & $1.14_{-0.04}^{+0.02}$       \\  
$\log(R_*/R_{\odot})$        & ${-1.68_{-0.15}^{+0.10}}^{(**)}$ & $-1.65 \pm 0.25$        & $-1.73_{-0.01}^{+0.025}$     \\  
$M_{\rm env}$ [$M_{\odot}$]  & ---                              & $(6-30) \times 10^{-7}$ & $0.019 \pm 0.006$            \\  
C/He, O/He$^{(*)}$           & $0.9, 0.4^{\rm (a)}$             &        ---              & $1.14, 0.71$                 \\   
BC [mag]                     & $-5.81_{-0.21}^{+0.23}$          &        ---              & $-5.89_{-0.04}^{+0.08}$      \\ 
$M_{\rm V}$ [mag]            & $7.55_{-0.51}^{+0.74}$           &        ---              & $7.79_{-0.10}^{+0.03}$       \\
$M_{\rm bol}$ [mag]          & $1.74$                           &        ---              & $1.9_{-0.14}^{+0.11}$        \\
$A_{\rm  V}$ [mag]           & $0.19$                           &        ---              & $0.071$ \\ 
$d$  [pc]                    & $682$                            & $700_{-400}^{+1000}$    & $614_{-32}^{+58}$            \\ 
$\pi$ [mas]                  & $1.47$                           &  $1.43_{-0.84}^{+1.9}$  & $1.6\pm0.1$                  \\ 
\hline
\hline
\end{tabular}
\label{tabla2}

{\footnotesize  Note: $(*)$ Abundances by mass, 
$(**)$ Interpolated from the tracks by assuming 
spectroscopic $(T_{\rm eff},\log g)= (80 \mbox{kK},7.5)$.}

{\footnotesize  References: (a)  Dreizler \& Heber (1998); 
(b) Miller Bertolami \& Althaus (2006).}
\end{table*}

\subsection{Characteristics of the best-fit model}
\label{char}

The main features of our best-fit model are summarized in Table
\ref{tabla2},  where we
also   include   the  parameters   of   \pg\   from  other   published
studies\footnote{Errors in $T_{\rm eff}$ and $\log(L_*/L_{\odot})$ are
estimated  from the  width of  the  maximum in  the function  $\chi^2$
vs. $T_{\rm  eff}$ and $\log(L_*/L_{\odot})$,  respectively; the error
in the stellar mass comes from the grid resolution in $M_*$. Errors in
the remainder  quantities are derived from these  values.}.  Note that
the effective temperature of our  best-fit model is virtually the same
as  the spectroscopic value.  Thus, the  location of  the star  in the
$\log  T_{\rm eff} -  \log g$  plane is  vertically shifted  to higher
gravities according to our predictions (see Fig. \ref{figure1}).

Our  best-fit model  has a  stellar  mass of  $M_*= 0.556  M_{\odot}$,
somewhat  smaller than  the  value  derived from  the  average of  the
computed period spacing, $M_*  \sim 0.57 M_{\odot}$, and substantially
lower than that inferred  from the asymptotic period spacing, $M_*\sim
0.63  M_{\odot}$ (see  \S \ref{period-spacing}).   On the  other hand,
FUEA07 have inferred  a value of the stellar mass of  \pg\ by using an
interpolation  formula to  the period  spacing derived  by  Kawaler \&
Bradley  (1994) on  the basis  of a  large grid  of  artificial PG1159
models  in  the  luminosity  range $1.6  \lesssim  \log(L_*/L_{\odot})
\lesssim 3.0$.  These authors obtain  a rather high value of $0.59 \pm
0.02  M_{\odot}$,  in line  with  the  trend  of early  determinations
(O'Brien  et al.  1998) and  also in  good agreement  with  our values
derived from the  period spacing, but in clear  conflict with the mass
of our best-fit model.

On the other  hand, the $M_*$ value of our  best-fit model is somewhat
higher  than the  spectroscopic mass  of $0.53  M_{\odot}$  derived by
Miller Bertolami \&  Althaus (2006) (see also Dreizler  \& Heber 1998)
for \pg.  Note that  a discrepancy between the asteroseismological and
the  spectroscopic  values of  $M_*$  is  generally encountered  among
PG1159 pulsators  (see C\'orsico et  al. 2006, 2007).  Until  now, the
asteroseismological  mass of  \pg\ has  been about  $10-30  \%$ larger
($\Delta  M_*  \approx 0.06-0.17  M_{\odot}$)  than the  spectroscopic
mass.   In light of  the best-fit  model derived  in this  paper, this
discrepancy is notably reduced to  less than about $5 \%$ ($\Delta M_*
\approx 0.026 M_{\odot}$).

FUEA07 infer  the stellar luminosity of  \pg\ by using  the formula of
Kawaler   \&   Bradley    (1994)   mentioned   above.    They   obtain
$\log(L_*/L_{\odot})= 1.3 \pm 0.5$,  larger than the luminosity of our
best-fit model,  $\log(L_*/L_{\odot})= 1.14_{-0.04}^{+0.02}$, and with
an accuracy a factor 20 worse. The large uncertainty in the luminosity
quoted   by  FUEA07   is  due   to  the   large  uncertainty   in  the
spectroscopically  determined  $\log  g$,  a quantity  used  by  these
authors to derive the luminosity.

\subsection{Helium-rich envelope thickness}
\label{helium}

An important parameter to be  discussed separately is the thickness of
the outer envelope ($M_{\rm env}$) of \pg.  We define $M_{\rm env}$ as
the mass above the chemical discontinuity between the He-rich envelope
and  the  C/O  core.   Our  best-fit model  has  $M_{\rm  env}=  0.019
M_{\odot}$.   On the  other hand,  FUEA07 suggest  a value  of $M_{\rm
env}$ in the range $(6 - 30) \times 10^{-7} M_{\odot}$, about 5 orders
of magnitude smaller. In this  section, we try to answer the question:
could a strikingly  low value of $M_{\rm env}$  like that suggested by
FUEA07 be  explained by  mass loss during  the PG1159 phase?   To this
end,  we  performed  additional  PG1159 evolutionary  calculations  to
explore  the amount of  stellar mass  that could  be eroded  by winds.
Specifically, we  have performed new evolutionary  simulations for the
sequence  of the  best-fit model  ($M_*= 0.556\,  M_{\odot}$) starting
from  the second  departure (post-VLTP)  of the  AGB until  the PG1159
stage  is  reached,  with  different  mass  loss  rate  prescriptions.
Specifically,  we   have  adopted   two  different  mass   loss  rates
($\dot{M}_1$,   $\dot{M}_2$)   appropriate   for  radiatively   driven
winds. Namely,  the one given by  Bl\"ocker (1995), which  is based on
Pauldrach et al. (1988), results

\begin{equation}
\dot{M}_1=1.29 \times 10^{-15} \, 
\left(\frac{L_*}{L_\odot}\right)^{1.86} \, [M_\odot/{\rm yr}],
\end{equation}

and the one adopted by Lawlor \& MacDonald (2006), which is based on a
modified version of the treatment of Abbott (1982),

\begin{equation}
\dot{M}_2=1.2 \times 10^{-15} \, \left(\frac{L_*}{L_\odot}\right)^2 \, 
\left(\frac{M_{\rm eff}}{M_\odot}\right)^{-1}\, 
\left(\frac{Z}{Z_\odot}\right)^{1/2} \, [M_\odot/{\rm yr}].
\label{m2}
\end{equation}

\noindent In  the last expression $M_{\rm  eff}=(1-\Gamma) M_*$ 
with $\Gamma$ defined as in Castor et al. (1975).  The metallicity was
set  to $Z=  Z_\odot$, because  at high  metallicities iron  lines are
expected to be dominant for  radiative driven winds (Vink et al. 2001)
and iron  abundance is expected  to remain unchanged during  the whole
evolution\footnote{However, how  important C,N,O  lines can be  at the
extremely high abundances of PG1159  stars is not known.  In any case,
we  think that  the  inclusion  of a  simulation  with $\dot{M}_3=  10
\dot{M}_2$ ---that would  correspond to the inclusion of  a value $Z /
Z_{\odot}=  100$, or  ``$Z= 2$''  in Eq.  (\ref{m2}); see  the text---
really sets  an upper  limit for possible  mass loss rates  during the
evolution.}.   Roughly, $\dot{M}_1$  is about  one order  of magnitude
lower than $\dot{M}_2$ in the present simulations. The total amount of
mass lost by  these sequences when they reach the  location of \pg\ is
$7
\times 10^{-5} M_\odot$ for $\dot{M}_1$ and
$4.4\times10^{-4}M_\odot$ for  $\dot{M}_2$, which are  both negligible
as compared with the mass of  the envelope of the best-fit model.  For
completeness we have considered a more extreme case by adopting a mass
loss rate of  $\dot{M}_3= 10\ \dot{M}_2$.  In this  case the mass loss
rate at the  WR-CSPN stage ($L_* \sim 10\,  000\, L_\odot$ and $T_{\rm
eff}< 100\,000$ K) is of the order of several $10^{-6} \, M_\odot/{\rm
yr}$,  and the rate  at the  evolutionary ``knee''  in the  HR diagram
during the  PG1159 stage  is of about  $10^{-7} \,  M_\odot/{\rm yr}$.
These values  are consistent  with the largest  rates observed  at both
PG1159  and WR-CSPN stages  (Koesterke et  al.  1998,  Koesterke 2001)
and,  consequently, are probably  an overestimation  of the  effect in
view of  the low mass  of our best-fit  model. Even in this  case, the
mass  eroded by  winds  amounts to  only $3.4\times10^{-3}\,  M_\odot$
ºwhich is about one order of  magnitude lower than the initial mass of
the envelope\footnote{It  is interesting to  note that, even  for this
extreme  case, the  period-fit does  not deteriorate  significantly as
compared    with   the    case    of   the    best-fit   model    (see
Sect. \ref{searching}).}.   Thus, it seems  that envelopes as  thin as
those proposed by  FUEA07 could be ruled out in  the context of single
star stellar evolution. More importantly, the reduction in the mass of the
He-rich envelope from a  canonical value of $\sim10^{-2} M_{\odot}$ to
a  value of  $\sim10^{-7}  M_{\odot}$ would  require  an extreme  fine
tuning (of  five orders of magnitude)  in the mass-loss  rate to avoid
the  complete removal  of the  whole envelope.   In the  absence  of a
mechanism  that  justifies  this  fine  tuning,  such  extremely  thin
envelopes should be taken with some caution.

\begin{figure}
\centering
\includegraphics[width=220pt]{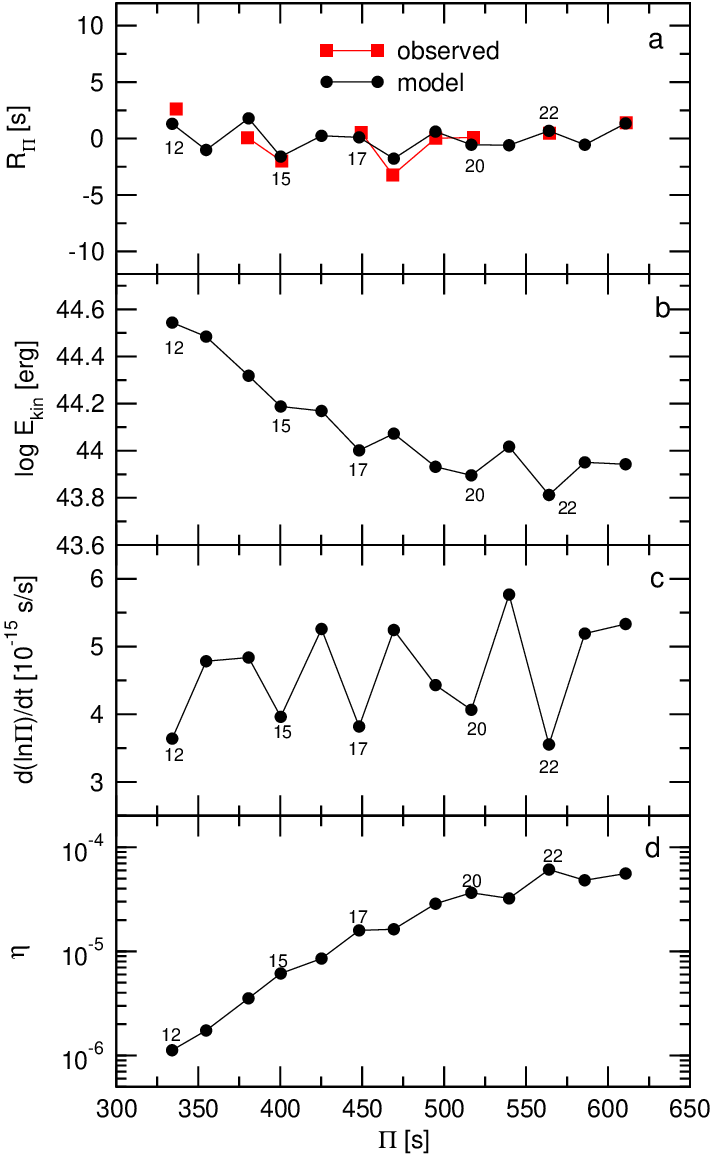}
\caption{Panel {\bf a}: Distribution of the residuals $R_{\Pi}$ relative 
to the mean period spacing for  the case of the observed periods (red)
and for  the case  of the calculated  periods (black) of  the best-fit
model.  Panel {\bf b}: the  distribution of the kinetic energy.  Panel
{\bf c}:  the values  of the relative  rates of period  change.  Panel
{\bf d}: the values  of the linear nonadiabatic stability coefficients
$\eta=   -\Im(\sigma)/  \Re(\sigma)$   (being  $\sigma$   the  complex
eigenfrequency). The numbers correspond to the radial order $k$ of the
modes trapped in the envelope.  See the text for details [Color figure
only available in the electronic version of the article].}
\label{figure5}
\end{figure}
 
\subsection{Mode trapping}
\label{mode-trapping}

In this section we shall try to disentangle the possible mode-trapping
signatures that could be hidden in the observed period spectrum of
\pg.  Following  FUEA07,  we  consider  the residuals ($R_{\Pi}$) of  
the    period    distribution   relative    to    the   mean    period
spacing\footnote{Residuals  relative to  the mean  period  spacing are
more appropriate than the forward period spacing ($\Delta \Pi_k=
\Pi_{k+1}- \Pi_k$) ---amply used in asteroseismology--- because of the
many missing  modes in the period  spectrum of \pg.}. For  the case of
\pg, a  linear least-square fit to  the observed periods  gives a mean
period spacing of $22.97$ s, while  for the best-fit model we obtain a
mean theoretical  period spacing  of $23.05$ s.   In panel {\bf  a} of
Fig.   \ref{figure5}  we   plot  the  $R_{\Pi}$-distribution  for  the
observed  periods (red)  and for  the case  of the  calculated periods
(black) of the best fit  model. The calculated distribution is in very
good agreement with the observed one, in particular for the modes with
$k= 15,  17, 19, 20,  22, 24$.  In  addition, the global  structure of
maxima  and  minima  seen  in  the  observed  distribution  is  nearly
duplicated by the computed one.

Mode trapping in PG1159 stars has been discussed at length by Kawaler
\& Bradley (1994) and C\'orsico \& Althaus (2006); we refer the reader
to those works for details. Here, we shall try to answer the question:
which modes  could be trapped in  the outer envelope of  \pg? At first
glance, they would be the modes showing a minimum in the $R_{\Pi}-\Pi$
diagram.  As we shall see below, this criterion can lead to erroneous
conclusions.  A  more secure way to  find which modes  are trapped in
the  outer  envelope is  to  examine  their  pulsation kinetic  energy
($E_{\rm kin}$).  In panel {\bf  b} of Fig.~\ref{figure5} we show the
kinetic energy distribution for  our best-fit 
model\footnote{ The kinetic energy values correspond to a 
normalization of the radial eigenfunction of $\xi_r/r= 1$ at the 
stellar surface.}.  Since modes that
oscillate  mainly  in the  outer  envelope  have  lower $E_{\rm  kin}$
values, one  can easily identify  trapped modes as those  having local
minima in  the kinetic energy distribution.   As can be  seen from the
figure, they are the modes with  $k= 12,15,17,20$ and 22. Note that in
some  cases a minimum  in $R_{\Pi}$  does coincide  with a  minimum in
$\log E_{\rm kin}$ (for instance for  $k= 15$) and in other cases does
not (for instance for $k= 17$).

Other useful quantities  to identify trapped modes are  the rates of
period changes ($\dot{\Pi}/\Pi$) and the linear stability coefficients
($\eta$). Modes trapped  in the envelope of the  model should ``feel''
more  strongly the  effects of  the surface  gravitational contraction
than untrapped modes,  and thus the former should  be characterized by
lower  values of  $\dot{\Pi}/\Pi$.   This is  clearly demonstrated  in
panel {\bf c}  of the figure, where we can see  that the trapped modes
are characterized  by local minima  in the distribution. On  the other
hand, it  is well known  from non-adiabatic arguments that  the linear
stability  coefficients are  larger for  modes characterized  by lower
kinetic energies.   This is depicted in  panel {\bf d}  of the figure,
where the  trapped modes (characterized by low  kinetic energies) have
local maxima in the $\eta$-distribution.

In view of the above discussion, since the mode at $\Pi \approx 468$ s
($k= 18$) ---which is identified as  a trapped mode by FUEA07--- has a
minimum in  the observed and computed  $R_{\Pi}$-distributions, but it
has a maximum in the kinetic energy, we conclude that this mode is not
a trapped mode in the outer envelope.  The mode at $\Pi \approx 401$ s
($k= 15$), on the other hand, corresponds to a minimum in the observed
and theoretical $R_{\Pi}$-distributions, and  a minimum of the kinetic
energy; so,  we conclude that  this is a  genuine trapped mode  in the
envelope, confirming the conclusion  of FUEA07.  However, the trapping
cycle of  about 68  s ($\Delta  k \approx 3$)  suggested by  FUEA07 is
unvalidated in  the frame  of the present  analysis since the  mode at
$\Pi \approx  468$~s  which is used  by those  authors would not  be a
trapped mode.

\begin{figure}
\centering
\includegraphics[clip,width=240pt]{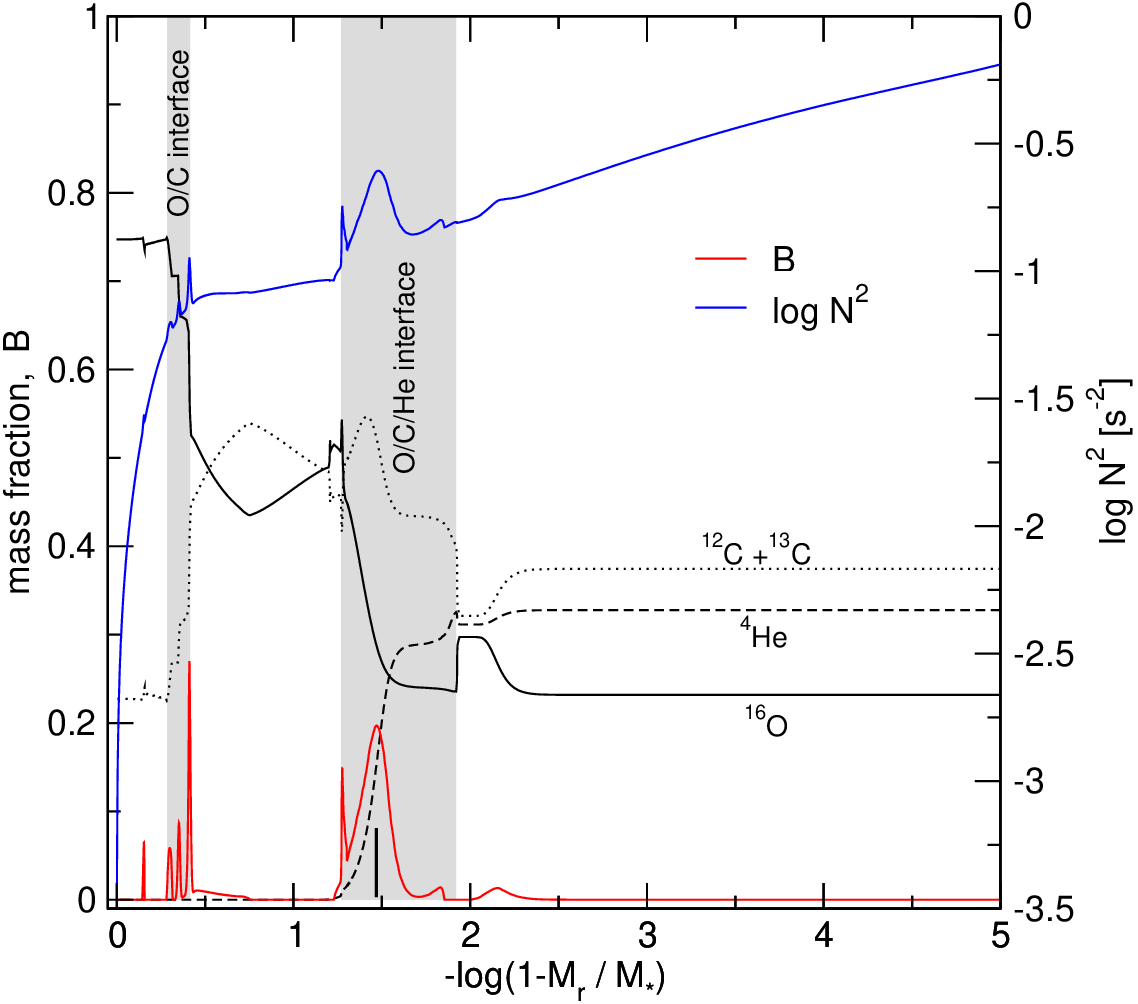}
\caption{The internal chemical profile of our best-fit 
model ($T_{\rm eff}=  81º,540$ K, $M_*= 0.556 M_{\odot}$)  in terms of
the outer fractional mass.  Also  shown are the profiles of the Ledoux
term $B$  and the logarithm  of the square of  the Brunt-V\"ais\"al\"a
frequency ($N$).  The thickness of the outer envelope is $M_{\rm env}=
0.024  M_{\odot}$.   The  location  of  the O/C  and  O/C/He  chemical
transition regions is emphasized  with gray regions. The vertical line
at $-\log(1-M_r/M_*)\sim 1.35$ marks the location of the bottom of the
envelope ($X_{\rm  He}\sim 0.14$) [Color figure only  available in the
electronic version of the article].}
\label{figure6} 
\end{figure}

In  closing, a  final  note  on the  mode-trapping  properties of  our
best-fit  model is  worth adding.  The variations  seen in  the period
distribution   ---as  revealed   by  the   $R_{\Pi}-\Pi$   diagram  of
Fig. \ref{figure5}---  are due  to mode-trapping effects  inflicted by
two chemical  transition regions: the  inner interface of O/C  and the
more external interface of  O/C/He.  The internal chemical profile and
the run of the Ledoux term $B$  and the logarithm of the square of the
Brunt-V\"ais\"al\"a frequency ($N$) of  our best-fit model in terms of
the outer fractional  mass are depicted in Fig.  \ref{figure6}. We can
wonder at this point whether  the O/C/He interface or the O/C chemical
transition region is more relevant at fixing the mode-trapping pattern
of our model, or if there exists a sort of core-envelope degeneracy in
the  sense that  both interfaces  are equally  effective  in producing
mode-trapping  structure (see Montgomery  et al.  2003). To  gain some
insight into  this direction, we have redone  our pulsation computations
by  minimizing the  influence of  a  given chemical  interface on  the
period  structure of  the best-fit  model\footnote{We employ  the same
procedure  like in  C\'orsico \&  Althaus (2005,  2006); we  refer the
reader to those papers for details.}.  We found that, at the domain of
the observed  range of periods  in \pg, the mode-trapping  features of
our model  are induced mostly by  the chemical gradient  at the O/C/He
interface,  being the O/C  interface much  less relevant.  For periods
longer  than about  $650-700$  s,  instead, it  is  the core  chemical
structure  in the  O/C interface that  mostly fixes  the  mode trapping
properties; this statement applies, for instance, to the cases of \pp\
and  \rxj\ (see  C\'orsico  \&  Althaus 2005,  2006  and C\'orsico  et
al. 2007 for more details).

\subsection{The asteroseismological distance and parallax of \pg}
\label{distance}

We employ  the luminosity of our  best-fit model to  infer the seismic
distance of \pg\ from the Earth. First, we consider the flux predicted
by a NLTE model atmosphere with $T_{\rm eff}= 80$ kK and $\log g= 7.5$
integrated through  the spectral response  of the V filter.  The model
atmosphere was calculated with the T\"ubingen Model Atmosphere Package
(see,  for  details, Werner  et  al.  2003).  We obtain  a  bolometric
correction BC= $-5.89$ and an absolute magnitude $M_{\rm v}= 7.79$. We
account  for  the  interstellar  absorption, $A_{\rm  V}$,  using  the
interstellar  extinction model developed  by Chen  et al.  (1998).  We
compute the seismic distance $d$ according to the well-known relation:
$\log d =  \frac{1}{5} \left[ m_{\rm v} - M_{\rm v}  +5 - A_{\rm V}(d)
\right]$ where  the apparent  magnitude is $m_{\rm  v}= 16.8  \pm 0.1$
(FUEA07).  The  interstellar  absorption  $A_{\rm  V}(d)$  varies  non
linearly with the  distance and also depends on  the Galactic latitude
($b$) and  longitude ($\ell$). For the equatorial  coordinates of \pg\
(Epoch  B2000.00,  $\alpha= 1^{\rm  h}\  25^{\rm  m}\ 22^{\rm  s}.00$,
$\delta=  +20^{\circ}\   17'\  54''.0$)  the   corresponding  Galactic
coordinates are  $b= -41^{\circ}\ 52'\ 1''.2$  and $\ell= 133^{\circ}\
38'\ 16''.8$. We solve for  $d$ and $A_{\rm V}$ iteratively and obtain
a  distance $d=  614_{-32}^{+58}$  pc and  an interstellar  extinction
$A_{\rm  V}= 0.0707$.   Note  that  our distance  is  $\approx 13  \%$
smaller than  the estimation of FUEA07  ($d= 700^{+1000}_{-400}$), and
with  its accuracy substantially  improved. Finally,  our calculations
predict a parallax of $\pi \sim 1.6$ mas.

In closing, we estimate a  ``spectroscopic'' distance of \pg. We first
derive $A_{\rm V}= 3.1\ E(B-V)=  0.19$ by employing the $E(B-V)= 0.06$
value from  Dreizler \& Heber  (1998). The distance can  be determined
from  the model  $V$-flux comparing  with  $m_{\rm V}$  and using  the
spectroscopic $T_{\rm  eff}$ and $\log  g$ and the  extinction $A_{\rm
V}$. We obtain a spectroscopic distance of $682$ pc, and a parallax of
$\sim  1.47$ mas.  We derive  also an  absolute magnitude  $M_{\rm v}=
7.55$  and  a  bolometric  correction  BC= $-5.81$  by  employing  the
``spectroscopic''  luminosity  and radius  ---  interpolated from  the
tracks assuming  spectroscopic $T_{\rm eff}$  and $\log g$---  and the
flux predicted by the model atmosphere (see Table \ref{tabla2}).

\section{Summary and conclusions}
\label{conclusions}

In this paper we carried  out an asteroseismological study of the cool
pulsating PG1159  star \pg, a  $g$-mode pulsator that defines  the red
edge  of  the GW  Vir  instability  domain  at low  luminosities.  Our
analysis is based on the full PG1159 evolutionary models of Althaus et
al.  (2005), Miller  Bertolami \& Althaus (2006) and  C\'orsico et al.
(2006).   These  models  represent   a  solid  basis  to  analyze  the
evolutionary and pulsational status of PG1159 stars like \pg.  This is
the second GW Vir star that  is pulsationally analyzed in the frame of
these  state-of-the-art PG1159  evolutionary models  ---the  first one
being  the  hottest  known  GW   Vir  star,  \rxj;  see  C\'orsico  et
al. (2007).

We first took advantage of the strong dependence of the period spacing
of variable  PG1159 stars  on the stellar  mass, and derived  a value
$M_* \sim 0.625 M_{\odot}$ by comparing $\Delta \Pi^{\rm O}$ with the
asymptotic period  spacing of our  models.  We also  compared $\Delta
\Pi^{\rm  O}$ with  the  computed period  spacing  averaged over  the
period range observed  in \pg, and derived a  value of $M_*\sim 0.567
M_{\odot}$. Note that in both derivations of the stellar mass we made
use of the spectroscopic constraint that the effective temperature of
the star should be $\sim 80$ kK. It is interesting to note that
the stellar mass as inferred from the asymptotic period spacing
is about $0.06 M_{\odot}$ larger than that derived from the 
average of the computed period spacings. This hints at possible 
systematics in the standard asteroseismological mass determinations methods, 
in particular when the full  asymptotic regime ($k \gg 1$) has
not been attained. We note that this systematics in the 
asteroseismological method is present not only in the case of full PG 1159 
evolutionary models as we use here, but also in PG 1159 models 
artificially created (see C\'orsico \& Althaus 2006). Because most analysis 
of pulsating PG1159 stars rely on the asymptotic 
period spacing, this point deserves to be explored for other
GW Vir stars, issue which we address  in a submitted paper.

Next, we adopted a less  conservative approach in which the individual
observed  pulsation  periods   alone  ---i.e.,  ignoring  ``external''
constraints such  as the spectroscopic  values of the  surface gravity
and     effective    temperature---     naturally    lead     to    an
``asteroseismological''   PG1159   model  that   is   assumed  to   be
representative of  the target star. Specifically,  the method consists
in looking  for the model  that best reproduces the  observed periods.
The period fit was  made on a grid of PG1159 models  with a quite fine
resolution in effective temperature ($\Delta T_{\rm eff}\sim 10-30$ K)
although admittedly  coarse in stellar  mass ($\Delta M_* \sim  0.01 -
0.08 M_{\odot}$). The match between  the periods of the best-fit model
and the observed periods in
\pg\ turns out be of an unprecedented quality for this type of studies, 
being  the   average  of  the  period   differences  (observed  versus
theoretical)  of only  $0.88$ s  with a  root-mean-square  residual of
$1.27$  s.   The  stellar  mass   of  the  best-fit  model   is  $M_*=
0.556_{-0.014}^{+0.009} M_{\odot}$.

Interestingly  enough,   the  mass   of  the  best-fit   model  ($M_*=
0.556_{-0.014}^{+0.009}  M_{\odot}$) is  closer  to the  spectroscopic
value  of $M_*= 0.53\pm  0.1 M_{\odot}$  (Dreizler  \&  Heber  1998; 
Miller Bertolami  \& Althaus 2006) than  the  asteroseismological  mass
derived in  previous works, of $0.59-0.69  M_{\odot}$ (FUEA07; O'Brien
et al. 1998).

Other characteristics of the best-fit model are summarized in Table
\ref{tabla2}. In particular, its effective temperature is nearly the 
same  (to within  $2 \%$)  as  the spectroscopic  $T_{\rm eff}$.   The
surface gravity, on the other  hand, is somewhat larger than the value
given by spectroscopy.  We also infer the ``seismic distance'' of
\pg\ by using the luminosity of our best-fit model. We obtain
a distance $d \sim 614$ pc, somewhat smaller than that of FUEA07.

Finally,  our computations predict  a temporal  period drift  for \pg\
between $1.22 \times 10^{-12}$ s/s and $3.26 \times 10^{-12}$ s/s. The
positive values  of $\dot{\Pi}$ (increasing periods)  reflect the fact
that our  best-fit model  is entering its  white dwarf  cooling domain
where  the  effect  of  the  increasing  electron  degeneracy  on  the
pulsation  periods  overwhelms   that  of  the  surface  gravitational
contraction, even for the modes trapped in the envelope. Strong  
theoretical arguments suggest that  \pg\ could be
used to constrain the plasmon  neutrino rates in the dense interior of
pre-white  dwarfs on  the basis  of an  observed value  of $\dot{\Pi}$
(O'Brien et al.  1998; O'Brien  \& Kawaler 2000).  We defer a thorough
exploration of this exciting issue to a forthcoming paper.

The  results of  the period-fit  procedure  carried out  in this  work
suggest  that  the  asteroseismological  mass  of  \pg\  ($\sim  0.556
M_{\odot}$) could be $\sim 6 - 20 \%$ lower than thought hitherto (see
O'Brien et al.  1998 and more recently FUEA07) and in closer agreement
(to within $\sim 5 \%$)  with the spectroscopic mass derived by Miller
Bertolami  \&  Althaus  (2006).    This  suggests  that  a  reasonable
consistency  between   the  mass  values  obtained   from  both  (very
different)  methods  should   be  expected  when  detailed  period-fit
procedures on  full PG1159 evolutionary models such  as those employed
in  this paper are  considered.  Even  more, \emph{a  better agreement
between asteroseismological  and spectroscopic masses of  GW Vir stars
could be found when  the same  evolutionary tracks  are used  for  both 
the asteroseismological and  the spectroscopic derivations  of the stellar
mass},  as we  do  in  the present  work\footnote{See  Quirion et  al.
(2007) for an enlightening note  about this topic.}. An anomalous case
in   this   context  could   be   \rxj,   for   which  we   found   an
asteroseismological   mass  about  $25   \%$  \emph{lower}   than  the
spectroscopic value  by employing the same  stellar evolution modeling
than here  (C\'orsico et al.  2007).   As we suggested  in that paper,
the  discrepancy  in  mass  could  be  due  to  large  errors  in  the
spectroscopic  determination of  $\log  g$ and  $T_{\rm  eff}$ for  RX
J2117.1+3412, and/or uncertainties in the location of the evolutionary
tracks in the HR  and $\log T_{\rm eff} - \log g$  diagrams due to the
modeling of PG1159 stars and their precursors\footnote{However, recent
work  by Miller Bertolami  \& Althaus  (2007b) suggests  that previous
evolution  does  not  play  a  crucial  role  in  shaping  the  PG1159
evolutionary tracks.}.
 
In any case, detailed asteroseismological period-fits for other GW Vir
stars based on  full evolutionary models like we  employ here, as well
as  precise spectroscopic determination  of the  effective temperature
and gravity of PG1159 stars will be needed in the future if we want to
reduce  the persisting  discrepancies  in the  stellar  mass of  these
fascinating stars.

In closing,  in this paper we have been able to find  a  PG1159 model  
that nicely reproduces the period spectrum observed in \pg\
\emph{without artificially tuning} the value of structural parameters such as
the thickness of the  outer envelope, the surface chemical abundances,
or the  shape of  the core chemical  profile which, instead,  are kept
fixed at the values predicted by our evolutionary computations. In some sense, 
this makes the fit derived more statistically significant. In particular,
our PG 1159 evolutionary models are characterized by thick helium-rich 
envelopes.  However, we cannot discard the possibility that pulsating  
PG1159 star could harbor thin helium-rich envelopes, a possibility sustained by 
the fact that PG 1159 and born-again stars are observed to suffer from appreciable 
mass loss. Resulting asterosesimological fits in this case would be worth 
exploring.


\begin{acknowledgements}
This paper has been benefited from a valuable referee report by
M. H. Montgomery. This research has been partially supported by the PIP 6521 grant 
from CONICET. This research has made use of NASA's Astrophysics Data System.
\end{acknowledgements}

\end{document}